\documentstyle[12pt, psfig, a4]{article}
\newcommand{\etal}{{\em et al.}}
\newcommand{\rsun}{R$_\odot$}

\begin{document}

\begin{center}

{\large\bf LASCO and EIT Observations of Coronal Mass Ejections}

\vspace{0.3cm}
K. P. Dere\\

Naval Research Laboratory, Code 7663, Washington DC 20375\\

\vspace{0.3cm}

A. Vourlidas, P. Subramanian\\

Center for Earth Observing and Space Research, Institute for
Computational Sciences, George Mason University, Fairfax, VA 22030\\

\end{center}

\begin{center}

Proceedings of the YOHKOH 8th Anniversary International Symposium,
``Explosive Phenomena in Solar and Space Plasma'', Dec 6-8 1999, Sagamihara,
Japan

\end{center}

The LASCO and EIT instruments on the SOHO spacecraft have provided an
unprecedented set of observations for studying the physics of coronal
mass ejections (CMEs).  They provide the ability to view the pre-event
corona, the initiation of the CME and its evolution from the surface of
the Sun through 30 \rsun.  An example of the capability of these
instruments is provided in a description of a single event (Dere \etal,
1997).  During the first 2 years of operation of LASCO and EIT on SOHO,
a substantial fraction, on the order of 25 to 50\%, of the CMEs
observed exhibited structure consistent with the ejection of a helical
magnetic flux rope.  An examples of these has been reported by Chen
\etal\ (1997)  and Dere \etal\ (1999).  These events may be the coronal
counterpart of magnetic clouds discussed by Burlaga \etal (1981) and
Klein and Burlaga (1982).  They analyzed observations of magnetic fields
behind interplanetary shocks and deduced that the field topology was
that of a helical flux rope.

Recently, we have explored a number of the consequences of the helical
flux rope description of these types of CMEs.  Vourlidas et al. (1999)
examined the energetics of CMEs with data from the LASCO coronagraphs
on SOHO. The LASCO observations provide fairly direct measurements of
the mass, velocity and dimensions of CMEs. Using these basic
measurements, we determined the potential and kinetic energies and
their evolution for several CMEs that exhibited a flux-rope morphology.
Assuming magnetic flux conservation ('frozen-in' fields), we used
observations of the magnetic flux in a variety of magnetic clouds near
the Earth to determine the magnetic flux and magnetic energy in CMEs
near the Sun.  Figure 1 shows these quantities for a few representative
flux rope CMEs. In general, we find that the potential and kinetic
energies increase at the expense of the magnetic energy as the CME
moves out, keeping the total energy roughly constant. This demonstrates
that flux rope CMEs are magnetically driven. Furthermore, since their
total energy is constant, the flux rope parts of the CMEs can be
considered to be a closed system above $\sim$ 2 $R_{\odot}$.

\begin{figure}
\psfig{file=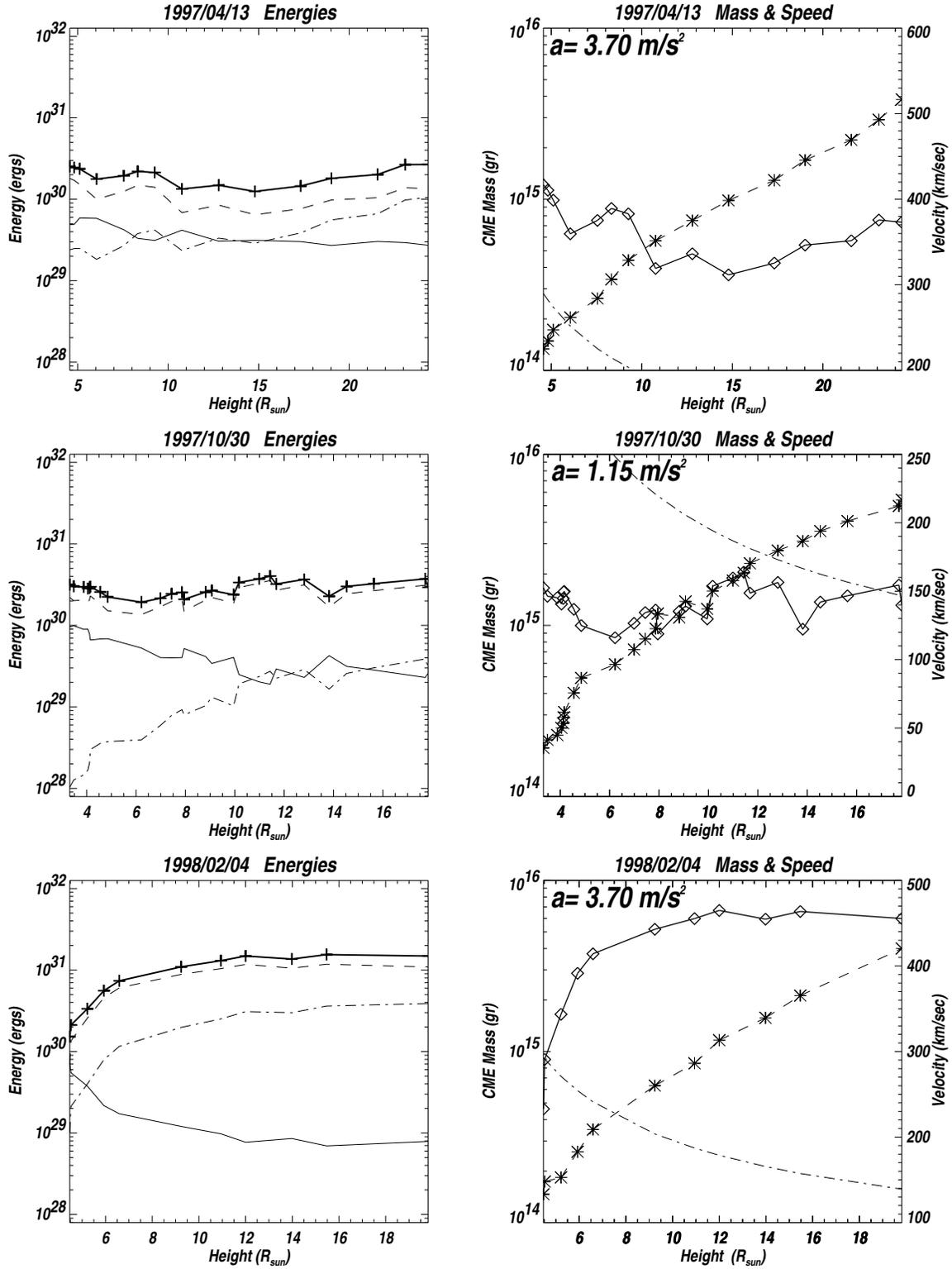,width=6.in}
\caption{On the left, the total (heavy line), potential (dashed line),
kinetic (dot-dash line) and magnetic (full line) energies of three
CMEs.  On the right, their mass (diamonds) and velocity (asterisks).}
\end{figure}

Subramanian et al. (1999) examined images from LASCO to study the
relationship of coronal mass ejections (CMEs) to coronal streamers.  We
wished to test the suggestion of Low (1996) that CMEs arise from flux
ropes embedded in streamers near their base.  It is expected that the
CME eruption would lead to the disruption of the streamer.  To date,
this is the most extensive observational study of the relation between
CMEs and streamers.  The data span a period of 2 years near sunspot
minimum through a period of increased activity as sunspot numbers
increased.  We have used LASCO C2 coronagraph data which records
Thomson scattered white light from coronal electrons at heights between
1.5 and 6$R_s$.  Synoptic maps of the coronal streamers have been
constructed from C2 observations at a height of 2.5$R_s$ at the east
and west limbs. We have superposed the corresponding positions of CMEs
observed with the C2 coronagraph onto the synoptic maps.  We identified
the different kinds of signatures CMEs leave on the streamer structure
at this height (2.5$R_s$). We find four categories of CMEs with respect
to their effect on streamers:
\begin{enumerate}
 \item CMEs that disrupt the streamer.\\[-0.8cm]
 \item CMEs that have no effect on the streamer, even though they are
related to it.\\[-0.8cm]
 \item CMEs that create streamer-like structures.\\[-0.8cm]
 \item CMEs that are latitudinally displaced from the streamer.\\[-0.8cm]
\end{enumerate}

Figure 2 summarizes these results.  CMEs in categories 3 and 4 are
not related to the streamer structure.  We therefore conclude that
approximately 35\% of the observed CMEs bear no relation to the
pre-existing streamer, while 46\% have no effect on the observed
streamer, even though they appear to be related to it.

Previous studies using SMM data (Hundhausen 1993) have made the general
statement that CMEs are mostly associated with streamers and that they
frequently disrupt it. Our conclusions thus significantly alters the
prevalant paradigm about the relationship of CMEs to streamers.

\begin{figure}[htb]
\centerline{\psfig{file=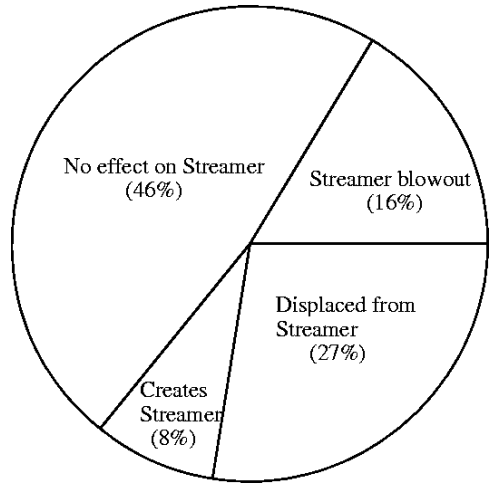,width=2.5in}} 
\caption{The relationship of CMEs to streamers}
\end{figure}

Subramanian and Dere (2000) have examined coronal transients observed
on the solar disk in EIT 195 \AA\ images that correspond to coronal mass
ejections observed by LASCO during the solar minimum phase of January
1996 through May 1998.  The objective of the study is to gain an
understanding of the source regions from which the CMEs observed in
LASCO images emanate.  We compare the CME source regions as discerned
from EIT 195 \AA\ images with photospheric magnetograms from the MDI on
SOHO and from NSO Kitt Peak, and also with BBSO H$\alpha$ images. The
overall results of our study suggest that a majority of the CME related
transients observed in EIT 195 \AA\ images are associated with active
regions.  We have carried out detailed case studies of 5 especially
well observed events.  These case studies suggest that active region
CMEs are often associated with the emergence of parasitic polarities
into fairly rapidly evolving active regions. CMEs associated with
prominence eruptions, on the other hand, are typically associated with
long lived active regions.  Figure 3 summarizes these results.

\begin{figure}[htb]
\centerline{
\psfig{file=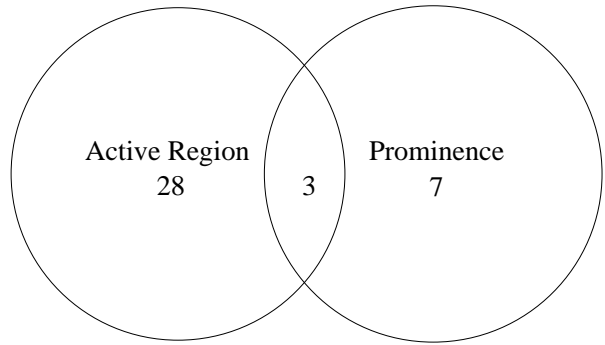,angle=270,width=3.1in}}
\caption{Coronal sources of CMEs}
\end{figure}

\begin{description}
\item[]\underline{References}
\item[]Chen, J., \etal\, 1997, ApJ, 338, L194\\[-0.8cm]
\item[]Dere, K. P., \etal, 1997, Solar Phys., 175, 601.\\[-0.8cm]
\item[]Dere, K. P., Brueckner, G. E., Howard, R. A., Michels, D.J.,
Delaboudiniere, J.P., 1999, ApJ, 516, 465.\\[-0.8cm]
\item[]Burlaga, L., Sittler, E., Mariani, F., Schwenn, R., 1981, JGR, 86, 6673\\[-0.8cm]
\item[]Klein, L. W., Burlaga, L. F., 1982, JGR, 87, 613\\[-0.8cm]
\item[]Subramanian, P. and Dere, K.P., 2000, ApJ, in preparation\\[-0.8cm]
\item[]Subramanian, P., Dere, K.P., Rich, N.B., Howard, R.A., 1999, JGR, 104, 22331\\[-0.8cm]
\item[]Vourlidas, A., Subramanian, P., Dere, K.P., Howard, R.A., 2000, ApJ, in press\\[-0.8cm]
\end{description}

\end{document}